# Sketching the Theory of Underdoped High Temperature Superconductors.

Moshe Dayan

Department of Physics, Ben-Gurion University,
Beer-Sheva 84105, Israel.

E-mail: mdayan@bgu.ac.il

# Sketching the Theory of Underdoped High Temperature Superconductors.

Moshe Dayan

Department of Physics, Ben-Gurion University,
Beer-Sheva 84105, Israel.

Abstract

The role of the anti-bonding state in the electron correlations in Copper oxide HTSC is analyzed. Then the t-J Hamiltonian is used to establish the formation of the charge stripes in underdoped oxides. It is proposed that these stripes make up the boundaries between the two degenerate anti-ferromagnetic (AFM) states, and that they are a key factor in switching between these states. We also provide a theoretical expression to the charge driven AFM magnons that have been observed by Neutron scattering experiments. Finally, the double correlation theory is applied to the stripe phase of holes to result in the superconductive gap and in the "pseudogap".

## 1. INTRODUCTION

In recent years I have developed a theory of the normal and the superconductive states of oxide HTSC [1-4]. The theory assumes large nesting sections in the Fermi surfaces of these materials. This nesting causes double-correlations, between members of Cooper-pairs, and electron-hole pairs. Consequently, it yields two gaps, the superconductive, as well as the correlation gap. The theory also yields CDW or SDW, depending upon the phase difference that one assumes between the waves of the two spin states. Another feature of the theory is the significant contribution of the Hartree diagram to the two order parameters. Retrospectively, it turns out that the contribution of the Hartree diagram applies only when the internal field is in a form of CDW, and not of SDW. In the undoped Copper oxides the internal fields are in forms of SDW. Thus, the treatment in [3, 4], taken as is, might apply more directly to non-magnetic HTSC, such as BKBO. As for the doped Copper Oxide HTSC, it is shown in the present work that the doping holes, in the under-doping range, agglomerate to form charge stripes which make the application of the said theory relevant. For the optimally-doping and the over-doping ranges the situation is still unclear, and additional investigation is needed.

There are two flaws in the discussed theory, which are closely related. The first is the lake of adequate treatment of doping. The second is that the nesting sections are treated only in general terms. Although the treatment does not limit the generality, it implicitly assumes that the nesting sections are in the nodal directions of the $d_{x^2-y^2}$ state of the Copper. The present reanalysis confirms the nesting and the instability in the nodal directions for the undoped oxides. However, for underdoped HTSC, the treatment needs to be readjusted, in view of the rich experimental data that suggest that both the pseudogap and the superconducting gap are directed towards the antinodes (A feature that is usually referred to as d-wave superconductivity) [5-8]. For the underdoped Copper oxide HTSC, the doping holes may be considered as excitations of the host anti-ferromagnetic (AFM) insulators. It is shown that these holes regroup to form stripes, and consequently, make up Fermi surfaces with nesting in the anti-nodal directions.

The present analysis utilizes the theory in [2-4] only after establishing the different background as outlined above. First we establish the role of the anti-bonding state in each cell of the un-doped parent insulators. Then, we show how neighboring cells bond to result in an AFM insulator. This state has qualitatively the same features as in [3], but differs substantially in the size of the energy gap. We show that the difference stems from the character of the anti-bonding band, where the Fermi level is located. After establishing the state of the parent undoped material, we build in it the Fermi system of the holes. We show that they tend to form linear conglomerates in the anti-nodal directions. This feature turns out to be in accord with the experimental Fermi surfaces, and in particular with their nesting directions. We show that these charge stripes result in magnetic collective excitations, which are in agreement with the magnetic satellites that have been inspected in neutron-scattering experiments. Then we apply the former theory, and compare the results with experiment. Due to the complexity of the doped copper-oxide HTSC, some parts of the present paper are only partially quantitative. Nevertheless, I believe that the scenario that is described in the following contains most of the main essential features of the theory of oxide HTSC.

## 2. THE ANTIBONDING STATE AND ITS IMPLICATIONS.

Band structure calculations on Copper-oxide HTSC were performed soon after the experimental discovery [9, 10]. All these early works were in general agreement. The valence band was found to be two dimensional in character, due to the large difference between the long Cu-O bonds in the z direction, and the relatively short Cu-O bonds in the x and y directions. The bonds in the x and y directions make $\sigma^*$ anti-bonding states between the $d_{x^2-y^2}$ state of the Copper, and the two $p_x$ and two $p_y$ states of the Oxygens. For one electron per bond, the $\sigma^*$ antibonding states are known to have molecular lobes along the lines that connect the bonded atoms, to have minimal probability amplitudes in some point between the atoms, and maximal probability amplitudes closer to the atomic cores. The Fermi surfaces, of the undoped parent materials, exhibit perfect nesting in the nodal directions of the $Cu-d_{x^2-y^2}$ state, with the wavevectors: $a^{-1}(\pm\pi,\pm\pi)$.

Let us start with a brief review of tight-binding approximation in a two dimensional cubic lattice. We start with only one atom per unit cell. In this approximation the Wannier functions may be replaced by the relevant atomic functions, to result in the Bloch state

$$\psi(k,r) = \frac{1}{\sqrt{N}} \sum_n \varphi(\mathbf{r} - \mathbf{R}_n) \exp(i\mathbf{k} \cdot \mathbf{R}_n) \tag{1}$$

where $\varphi(\mathbf{r} - \mathbf{R}_n)$ is the atomic function of the atom at the lattice vector $\mathbf{R}_n$. It is the eigenfunction of the atomic Hamiltonian, with the atomic potential $V^{at}(\mathbf{r} - \mathbf{R}_n)$, and with the eigenvalue $E^{at}$. As a function of $(\mathbf{r} - \mathbf{R}_n)$, it has the periodicity of the lattice. When $\psi(k,r)$ is applied to the crystalline Hamiltonian, and is multiplied with $\varphi^*(r)$, and integrated, we get

$$[\varepsilon_k - E^{at}] \sum_n \exp(i\mathbf{k} \cdot \mathbf{R}_n) \int \varphi^*(\mathbf{r}) \varphi(\mathbf{r} - \mathbf{R}_n) d\tau$$

$$= \sum_n \exp(i\mathbf{k} \cdot \mathbf{R}_n) \int \varphi^*(\mathbf{r}) [V(\mathbf{r}) - V^{at}(\mathbf{r} - \mathbf{R}_n)] \varphi(\mathbf{r} - \mathbf{R}_n) d\tau \tag{2}$$

where $\varepsilon_k$ is the band energy, and $V(\mathbf{r})$ is the actual potential. On the left hand side of Eq. (2), the dominant term correspond to n=0, whereas on the right hand side, apart from the n=0 term, the dominant contribution comes from the nearest neighbors. Thus we get

$$\varepsilon_k = E^{at} + A - 2t[\cos(k_x a) + \cos(k_y a)] , \tag{3}$$

with $A = \int \varphi^*(\mathbf{r})[V(\mathbf{r}) - V^{at}(\mathbf{r})]\varphi(\mathbf{r})d\tau$, $t = -\int \varphi^*(\mathbf{r})[V(\mathbf{r}) - V^{at}(\mathbf{r} - a\mathbf{i})]\varphi(\mathbf{r} - a\mathbf{i})d\tau$, where we have assumed an equal t for the nearest neighbors in the x and in the y-directions. When one considers a material with one electron per atom, the Fermi surface obtained from Eq. (3) is the well-known "diamond"-like surface that is made of the four straight lines connecting between the four points $(0, \pm\pi/a)$, and $(\pm\pi/a, 0)$, each to its two neighbors.

The former paragraph, with the assumption of one atom per unit cell, is an introductory preview. The unit cell of Copper oxide HTSC includes one Cu atom at its center, and four Oxygen atoms, each shared with the neighboring cell. The $Cu - d_{x^2-y^2}$ is anti-bonded to the O-p$_x$, and to the O-p$_y$. There are five possible states for a single electron per cell. Consequently, in analogy with the molecular bonding and antibonding problem, the eigenfunction is the vector $\Psi = \exp(-i\varepsilon_+ t/\hbar)[a_d \quad a_x \quad a_{-x} \quad a_y \quad a_{-y}]$, and the Hamiltonian is

$$H_{ab} = \begin{bmatrix} \varepsilon_d & V_{dp}\exp(-ik_x a/2) & V_{dp}\exp(ik_x a/2) & V_{dp}\exp(-ik_y a/2) & V_{dp}\exp(ik_y a/2) \\ V_{dp}^*\exp(ik_x a/2) & \varepsilon_p & 0 & 0 & 0 \\ V_{dp}^*\exp(-ik_x a/2) & 0 & \varepsilon_p & 0 & 0 \\ V_{dp}^*\exp(ik_y a/2) & 0 & 0 & \varepsilon_p & 0 \\ V_{dp}^*\exp(-ik_y a/2) & 0 & 0 & 0 & \varepsilon_p \end{bmatrix}$$

(4)

where $\varepsilon_d$ and $\varepsilon_p$ are atomic energies, and $V_{dp}$ is the hoping potential. The exponentials in the first row and column are due to the crystalline periodicity. Actually, they should not appear in this stage but only when the Bloch function is constructed. Here we show that even if introduced in this early stage, they are eliminated from the anti-bonding energy. Solving the Schrodinger equation requires the vanishing of the determinant of the characteristic matrix, namely $|H_{ab} - \varepsilon_\pm| = 0$. This yields

$$\varepsilon_\pm = \varepsilon \pm \sqrt{\varepsilon_{dp}^2 + 4|V_{dp}|^2} \quad . \tag{5}$$

In Eq. (5), $\varepsilon = (\varepsilon_d + \varepsilon_p)/2$, and $\varepsilon_{dp} = (\varepsilon_d - \varepsilon_p)/2$. From the ionization potentials of the oxygen and the $Cu - d_{x^2-y^2}$ we get, $\varepsilon = -16.95eV$, and $\varepsilon_{dp} = -3.35eV$. The two different solutions in Eq. (5) correspond to the bonding and antibonding states. The higher energy $\varepsilon_+$, corresponds to the antibonding state. In addition to the two solutions of Eq. (5), there are three other non-bonding solutions. Notice the lake of oscillating factors in Eq. (5). This is contrary to other works, which assumed three states problem, by performing initial summation of the two O-$p_x$ states, and the two O-$p_y$ states [10]. Consequently, in their results oscillating factors are present under the square-root sign, $\varepsilon_\pm = \varepsilon \pm \sqrt{\varepsilon_{dp}^2 + 4V_{dp}^2[\sin^2(\frac{k_x a}{2}) + \sin^2(\frac{k_y a}{2})]}$. Summing pairs of the probabilities prior to the formal solution includes an inherent assumption that the Cu-O hopping is the cell to cell crystalline hopping, whereas we have taken a different perception, in which the cell to cell hopping occurs between anti-bonding cellular molecules. The overlap integral for this hoping is a molecular overlap, whose contribution comes mostly from the linking Oxygen. The fact that $\varepsilon_\pm$ in Eq. (5) does not depend upon k is convenient, because it enables us to apply the one atom formalism of Eqs. (1-3) immediately to the multi-atom problem (per cell) under consideration. One has simply to replace $E^{at}$ of Eq. (3) with $\varepsilon_+$ of Eq. (5). The overlap integrals A and t may be calculated by means of the antibonding wave function which is calculated in the following paragraph.

The antibonding wave function is a linear combination of the $d_{x^2-y^2}$ function and the $p_x$, and $p_y$ functions: $\varphi_d = N_d \frac{x^2 - y^2}{r^2}$, $\varphi_x = \pm N_p \frac{x}{r}$, $\varphi_y = \pm N_p \frac{y}{r}$, where $N_d$ and $N_p$ are normalization numerical factors, and $r = \sqrt{x^2 + y^2}$. The signs of $\varphi_x$ and $\varphi_y$ are chosen according to the criterion of minimal amplitudes in the space between the atoms. The Schrodinger equation then yields

$$V_{dp}^* a_d = (\varepsilon_+ - \varepsilon_p)c \tag{6a}$$

$$4V_{dp} c = (\varepsilon_+ - \varepsilon_d)a_d \tag{6b}$$

with $c = a_x \exp(-ik_x a/2) = a_{-x} \exp(ik_x a/2) = a_y \exp(-ik_y a/2) = a_{-y} \exp(ik_y a/2)$.
Normalization implies that $|a_d|^2 + 4|c|^2 = 1$, which yields

$$|a_d|^2 = \frac{(\varepsilon_+ - \varepsilon_p)^2}{(\varepsilon_+ - \varepsilon_p)^2 + 4|V_{dp}|^2} \tag{7a}$$

$$|c|^2 = \frac{|V_{dp}|^2}{(\varepsilon_+ - \varepsilon_p)^2 + 4|V_{dp}|^2} \tag{7b}$$

The single electron antibonding wave function is a linear combination of atomic orbitals (LCAO)

$$\Psi_{ab}(\mathbf{r}) = N_d |a_d| \frac{x^2 - y^2}{x^2 + y^2} + N_p |c| \{ \frac{(x - \frac{a}{2})e^{ik_x a/2}}{\sqrt{y^2 + (x - \frac{a}{2})^2}} - \frac{(x + \frac{a}{2})e^{-ik_x a/2}}{\sqrt{y^2 + (x + \frac{a}{2})^2}}$$

$$- \frac{(y - \frac{a}{2})e^{ik_y a/2}}{\sqrt{x^2 + (y - \frac{a}{2})^2}} + \frac{(y + \frac{a}{2})e^{-ik_y a/2}}{\sqrt{x^2 + (y + \frac{a}{2})^2}} \}. \tag{8}$$

The signs of the p-orbitals have been fixed to cause minimum amplitudes at some points on the lines connecting the Oxygens and the Copper, as expected from an anti-bonding wave-function.

The anti-bonding function of Eq. (8) is bonded with its neighbors to form a band. Let us deal first with the case of one electron per site, $n = 1$. If a site is occupied by one electron, then a second electron of opposite spin can reside on the same site only when overcoming a large coulomb repulsion. This is controlled by the Hubbard Hamiltonian,

$$H_H = -t \sum_{<ij>s} c_{is}^+ c_{js} + U \sum_i n_{i\uparrow} n_{i\downarrow} \tag{9}$$

In Eq. (9), $<ij>$ indicates that the sum is taken only for nearest neighbors, $c_{js}$ and $c_{is}^+$ are annihilation and creation operators of the anti-bonding states (at sites j and i), respectively, t is the nearest neighbor hoping potential, and $n_{is}$ is the number of electrons in the spin state s and the site i, $n_{is} = 0,1$. The second term in Eq. (9) is a correlation term resulting from the Coulomb repulsion between two electrons that reside on the same site.

In $H_U$, we define $\delta n_{is} = n_{is} - <n_{is}>$ and obtain $H_U = H_{AF} + H_C$, where

$$H_{AF} = -t \sum_{<ij>s}(c_{is}^+ c_{js} + hc) + U \sum_{is} <n_{i-s}> n_{is} \qquad (10)$$

$$H_C = U \sum_i \delta n_{i\uparrow} \delta n_{i\downarrow} + E_0 \qquad (11)$$

and $E_0 = -U \sum_i <n_{i\uparrow}><n_{i\downarrow}>$. It is shown below that for the half-filled case (undoped Copper Oxides), the ground state of $H_{AF}$ is the AFM Néel state. In this ordered ground state, $H_C - E_0$ may be ignored, and $E_0$ is a constant. For $H_{AF}$, we assume two sub-lattices, A for spin up, and B for spin down, and make the ansatz

$$<n_{is}> = \frac{1}{2}[1 + 2 s_z m_0 \exp(-i\mathbf{G} \cdot \mathbf{R_i})]. \qquad (12)$$

In Eq. (12), $s_z = \pm \frac{1}{2}$, depending on the spin state, $\mathbf{G} = a^{-1}(\pi,\pi)$, $\mathbf{R_i}$ is the lattice vector for the site i. The quantity $m_0$ is the local magnetization. Inserting Eq. (12) into Eq. (10), and Furrier transforming yields

$$H_{AF} = \sum_{k<k_F,s}[(\varepsilon_k + \frac{U}{2})c_{ks}^+ c_{ks} + (-\varepsilon_k + \frac{U}{2})c_{k+G,s}^+ c_{k+G,s} - U s_z m_0 (c_{k+G,s}^+ c_{ks} + c_{ks}^+ c_{k+G,s})]. \qquad (13)$$

We shall see that the present problem is similar qualitatively to the problem that was analyzed in [3]. The difference is the size and the origin of $U$. Despite this difference, one may use the inherent analogy, which stems from the same symmetry. This makes us assume the ground state to be

$$|\Phi_0'> = \prod_{k<k_F,s}(v_k' c_{ks}^+ + 2 s_z u_k' c_{\bar{k},s}^+)|0> \qquad (14)$$

where $|0>$ is the vacuum, and the product is over k- states within the Fermi surface (of the uncondensed state). Applying the Hamiltonian to the ground state yields

$$H_{AF} |\Phi_0'> = \sum_{k<k_F,s} (\frac{U}{2} - E_k') |\Phi_0'> \qquad (15a)$$

$$E_k' = \sqrt{\varepsilon_k^2 + (m_0 U/2)^2} = \sqrt{\varepsilon_k^2 + \Lambda_k'^2} \qquad (15b)$$

The first equality in Eq. (15b) resulted from a direct calculation, whereas the second is a definition of $\Lambda_k'$, in accordance with [3]. Notice that $\Lambda_k' = 2s_z u_k' v_k' E_k'$ - is dependent upon k, whereas $m_0$ is not. This stems from the different treatments between the present work and Ref. [3]. This suggests that the magnetization is related to $\Lambda_k'$ by- $m_0 U/2 = \sum_k \Lambda_k'/N_0$, where $N_0$ is the number of occupied k- states.

The Hamiltonian $H_{AF}$ may be diagonalized by the transformation

$$\gamma_{ks} = -2s_z u_k' c_{ks} + v_k' c_{\bar{k},s} \qquad (16a)$$

$$\eta_{ks}^+ = v_k' c_{ks} + 2s_z u_k' c_{\bar{k},s} \qquad (16b)$$

The Hamiltonian then becomes

$$H_{AF} = \sum_{k<k_F,s} [(E_k' - \frac{U}{2})\eta_{ks}^+ \eta_{ks} + (E_k' + \frac{U}{2})\gamma_{ks}^+ \gamma_{ks} + (\frac{U}{2} - E_k')] \qquad (17)$$

The excitations $\gamma_{ks}^+$ and $\eta_{ks}^+$ may be considered as two types of excitations, the first is of particles above the energy gap, whereas the second is of anti-particles whose excitation leaves voids below the energy gap. If the Fermi energy is artificially shifted up by $U/2$, then the same quantity may be eliminated from both of the excitation energies.

Thus, we have shown that the nesting feature of Eq. (3) makes the one electron per-site metallic system unstable against the formation of an AFM Néel state, even for small size of correlation potentials. The size of the energy gap is an outcome of the anti-bonding character of the band. We have also demonstrated that the problem may be handled by means of the Hubbard Hamiltonian, despite the lack of strong Coulomb on-site repulsion.

### 3.  THE DOPED COPPER-OXIDE HTSC.

The solution of the last section for undoped oxides portrays an AFM Néel state. This includes two degenerate states, one in which the sublattice A includes only spin-up electrons, and the sublattice B only spin-down electrons. In the other state the spin states interchange locations. We know of many two state systems that oscillate between the two states while reducing the system energy. The question for the present case is: What is the most likely oscillation mechanism, and what are its effects? Answering this question is attempted in the present paper. In this respect the situation is clearer in doped oxides than in the undoped materials. This is so mainly because of the larger amounts of experimental data that exist for the doped materials. In the present analysis we shall deal mostly with underdoped Copper oxides, which are doped with holes. There is a large amount of experimental evidence that, in these materials, the holes aggregate to form rows or columns in the Cu-O bond directions. Here we claim that these charged stripes also make-up the domain boundaries between the discussed two degenerate states, which will be denoted as states C (where sublattice A accommodates spin-up electrons) and D (where sublattice B accommodates spin-up electrons).

When $t/U \ll 1$, one may project into the **reduced Hilbert space in which doubly occupied sites are excluded**, and transform the Hubbard Hamiltonian (9) into an approximated t-J Hamiltonian [11,12]

$$H_{tJ} = -t\sum_{\langle ij\rangle s}(c_{is}^+ c_{js} + hc) + J\sum_{\langle ij\rangle}(\mathbf{S}_i \cdot \mathbf{S}_j - \frac{n_i n_j}{4}) \ . \tag{18}$$

In Eq. (18), $J = 4t^2/U < t$, $S_i = \frac{1}{2}\sum_{\alpha\beta} c^+_{i\alpha}\sigma_{\alpha\beta}c_{i\beta}$, where $\sigma$ is the vector of the Pauli matrices, and $n_i = n_{i\uparrow} + n_{i\downarrow}$. Eq. (18) is in accord with experimental results in the underdoped regime. Besides, the t-J Hamiltonian is easier than the Hubbard Hamiltonian for getting some physical insights in the underdoped oxide superconductors, and consequently, it will be used in some of the following discussions. The quasi-particle excitations $\gamma^+_{ks}$, and $\eta^+_{ks}$ of Eqs.(16) are assumed to have momenta close to the diamond-like Fermi surface of the undoped material. This is so only because their correlations were not taken into account. Eq. (18) shows that when holes are injected into the material, an energy reduction is achieved when they agglomerate. This is so because two separated lone holes raise the energy of the second term of the Hamiltonian by $4J$ (relative to the AFM undoped material), whereas they raise it only by $3.5J$ if agglomerated into an attached pair. When n holes agglomerate to form a linear straight chain, the J energy is reduced from $2nJ$ to $(1.5n + 0.5)J$. When n holes agglomerate to form a squared domain, the J energy is reduced even more, to $(n + \sqrt{n})J$. However, this simple estimation scheme does not take account of the Coulomb energy, which does not favor this last configuration. Additional reason in favor of the linear holes grouping is that the stripe configuration is suitable to produce the mechanism that switches between the AFM states C and D, with the resultant additional energy reduction. The kinetic energy term (the t-term) does not violate this simple energy estimate, and even reinforces it. For lone holes close to the "diamond-like" Fermi surface, the energy $-2t[\cos(k_x a) + \cos(k_y a)]$ is close to zero, because the sum of the cosines at the Fermi surface is zero. For conglomerated holes the kinetic term is negative, which means that it favors agglomeration, too. Experimental results (especially of neutron scattering) suggest that the holes agglomerate to form domains of straight linear stripes in the x and y directions [13-16]. Fig. 1a depicts a model of several striped domains. Fig. 1b depicts how columns of holes make the boundaries between adjacent states C and D. In the following discussion we shall show that the eigenstates of charge stripes are made of linear combinations of the states that are depicted in the figure.

Let us describe quantitatively one of the rows of the state that is depicted in Fig. 1b,

$$\Phi_l^m = c_{m,1\downarrow}^+ \cdot c_{m,2\uparrow}^+ \cdots c_{m,l-2\downarrow}^+ \cdot c_{m,l-1\uparrow}^+ \cdot 1 \cdot c_{m,l+1\downarrow}^+ \cdot c_{m,l+2\uparrow}^+ \cdots c_{m,2N\downarrow}^+ = \Phi_C^m \cdot \Phi_D^m \qquad (19)$$

In Eq. (19), m is a row index, $1 \le m \le 2N$, where $2N$ is the number of lattice sites in the x and in the y directions. The figure 1 was put in the *l*-th site, to indicate the location of the hole. The choice of an equal number of rows and columns is only a mater of convenience, to match the simple configuration in Fig. 1a. The functions $\Phi_l^m$ are not eigenstates of the t-J Hamiltonian, neither is the linear combination

$$\Phi_q = (2N)^{-1} \sum_{l,m=1}^{2N} \Phi_l^m \exp[ia(q_x l + q_y m)]. \qquad (20)$$

However, it could be shown that $\Phi_q$ is the eigenfunction of $H_{tC}$, where $H_{tC} \propto H_t U_C$, $H_t$ is the hopping part of the t-*J* Hamiltonian, and $U_C$ is the time development operator that move the column of holes

$$H_{tC} \Phi_q = 2J \cos(qa) \Phi_q \qquad (21)$$

From the former discussion it is clear that the problem dimensionality has been reduced to one. When the stripe is along the y direction (a column stripe), the wavevector is in the x direction, and the eigenvalue of Eq. (21) is independent of y, and wise-versa for a row stripe. Although $\Phi_q$ is a collective excitation, it could be referred to as a collection of Fermions, which fill up a kind of Fermi area in a squared reciprocal zone. The borders of this area make-up some kind of a Fermi surface. Let us examine, for example, a squared domain of $2N \times 2N$ lattice sites with column stripes. The possible wavevectors are in the x-direction. The smallest non-trivial q that are in accordance with the cyclic boundary conditions are $\pm q_{mn} = \pm a^{-1}\pi(N^{-1},0)$. The largest q within the Fermi area are determined by the fractional hole concentration $\delta$ as $\pm q_\delta = \pm a^{-1}\pi(\delta,0)$. These $\pm q_\delta$ fit exactly the satellite wavevectors that were observed by Neutron scattering experiments [13-16]. The zone, however, goes beyond $\pm q_\delta$ of our underdoped samples. The last state by the boundary of the zone matches a wave that is made of only two AFM **electronic columns** in the domain, whereas the rest is full of holes. This is the case of $\delta_{mx} = 1 - N^{-1} \approx 1$, with $q_{mx} = a^{-1}\pi(\delta_{mx},0) \approx a^{-1}\pi(1,0)$. The ground state of the t-J Hamiltonian is

$$\Phi_\delta = \prod_{q_x=-q_\delta}^{q_\delta} \Phi_{qx} .$$

Now we wish to define the field of Fermions (holes) that drive the magnon which is expressed by $\Phi_{qx}$. Naturally, we work with holes operators. Notice that the two spin states are equally involved with $\Phi_q$. Therefore, we define $b_q^+ = (c_{q\uparrow} + c_{q\downarrow})/\sqrt{2}$, and the t-J field as $\Psi_{tJ}(x) = \sum_q \Phi_q(x) b_q$. The one dimensionality of the field allows us to work with $\Phi_{qx}$, even when we seek to express the x dependence of $\Phi_q$. We express $\Phi_{qx}$ as a combination of $\Phi_l$, where the latter is expressed by means of $\Phi_C$ and $\Phi_D$, which are restricted to their areas by means of step functions $\theta[\pm(x-la)]$

$$\Phi_l = \Phi_C \theta(la-x) + \Phi_D \theta(x-la) \tag{24}$$

$$\Phi_{qx} = (2N)^{-1/2} \sum_{l=1}^{2N} \exp(iq_x la) \Phi_l \tag{25}$$

Eq. (14) suggests that

$$\Phi_{C,D} = \prod_{k<k_F s} (v_k' c_{ks}^+ \pm 2s_z u_k' c_{\bar{k},s}^+) |0> \tag{26}$$

We use the integral expression of the step functions, $\theta(\pm x) = \mp \int \frac{dk' \exp(-ik'x)}{(2\pi)(k' \pm i\delta)}$. The function $\Phi_D$ has an even part, which continuous smoothly the function $\Phi_C$ through the boundary $x = la$, and an odd part which reverses the sign of the step function, and results in $\Phi^{odd}$. Only the second part contributes to $\Phi_q(x)$, and gives

$$\Phi_q(x) = \frac{1}{q_x} \exp(iq_x x) \Phi^{odd}(x) . \tag{27}$$

The function $\Phi^{odd}(x)$ is not expressed explicitly. It is a sum of products of terms such as $2s_{2z}v'_{\vec{k}1}u'_{\vec{k}2}c^+_{\vec{k}1s1}c^+_{\vec{k}2s2}|0>$, which are an expression of the AFM SDW with the wavevector $\mathbf{G}=a^{-1}(\pi,\pi)$. The exponential term modulates this SDW in the x-direction, by the smaller wavevector- $q_x$.

Our choice to work with $\Phi_{qx}$, rather than with $\Phi_q$, stems from the fact that the eigenvalue of $\Phi_q$ does not include the exchange energy between neighboring rows. This is disturbing when one wishes to define a Fermionic field, whose natural states should be the single Fermionic function $\Phi_q$. A simple way around this difficulty would be to replace $H_{tJ}$ by an effective t-J Hamiltonian, in which the exchange energy between neighboring rows is included in the exchange energy within the single row. This is done simply by not counting nearest neighbors of the said row in the sum for the exchange, and compensating for this by adding the neglected part to the exchange energy within the row. The said effective t-J Hamiltonian then should be

$$\tilde{H}_{tJ} \cong \{\sum_{q_x,s}\frac{1}{2}\varepsilon_{qx}c_{qs}c^+_{qs}\} - 2J(N-1) \tag{28}$$

The eigenvalue of $\Phi_q$ then should be $\tilde{H}_{tJ}\Phi_q \cong [\varepsilon_{qx} - 2J(N-1)]\Phi_q = \tilde{\varepsilon}_{qx}\Phi_q$. One should notice, however, **that $\tilde{H}_{tJ}$ is adequate only for the states under consideration, but not for any state in general**.

### 4. THE DOUBLE CORRELATED STATE

So far we have dealt with the undoped AFM Néel insulator, and its underdoped version. It is clear from the last two sections that both are condensed phases. However, the condensation of the underdoped material has not yet been clarified. The conglomeration of the doping holes to linear stripes has been justified by reducing the exchange energy between neighboring holes. We have also argued that these charge stripes make the boundaries between the two neighboring AFM states C and D, and that their movement acts as a mechanism for switching between the two states. This is a two degenerate states system. These systems are known to switch between the two states, while causing an energy reduction in the condensed phase. In the present section we discuss the formal theory of this condensed phase.

We propose that the field $\Psi_{tJ}(x) = \sum_q \Phi_q(x) b_q$ undergoes additional condensation similar to the double condensation discussed in [4]. All the background for establishing such a theory has already been set. The following discussion is only a brief review, and the detailed and rigorous derivation goes along the lines of [4]. The various quantities used in the following derivation are also denoted as in [4], unless stated otherwise. We use the operators $b_q$, but when it is necessary, we use $c_{qs}$. Notice that the zero of the $\varepsilon_k$ energy parameter in Ref. [4] was defined to be at the Fermi energy of the uncondensed phase. We use the same reference level in the following. Thus, in the following equations the zero of $\tilde{\varepsilon}_{qx}$ should also be at the Fermi level of the uncondensed phase. The discussed theory is applied with respect to the nesting that occurs in the q-space between the two momenta $\pm q_\delta = \pm a^{-1} \pi(\delta, 0)$. An equivalent situation occurs in domains of horizontal stripes, where $\pm q_\delta = \pm a^{-1} \pi(0, \delta)$, but in the following we keep dealing with the example of column stripes. Fig. (2) depicts the q-space with respect to the k-space. The figure also demonstrates the momenta of the satellites that were observed in Neutron scattering experiments [13-16]. At zero temperature, and without condensation, all states within the Fermi area are full, and states out of that area are empty. However, in a similar way to the superconductive state, there is a certain section in the q-space at which the states are spread in order to maintain the condensation. Let us define $\bar{q} = q - sign(q) 2|q_\delta|$. The 4-dimentional Nambu-like field of the condensed phase is

$$\tilde{\Psi}_q = \frac{1}{\sqrt{2}} \{ \begin{pmatrix} c^+_{qs} \\ c_{-\bar{q},-s} \\ c^+_{\bar{q}s} \\ c_{-q,-s} \end{pmatrix} + \begin{pmatrix} c^+_{q,-s} \\ c_{-\bar{q},s} \\ c^+_{\bar{q},-s} \\ c_{-q,s} \end{pmatrix} \} = \frac{1}{\sqrt{2}} \{ \tilde{\Psi}_{q,s} + \tilde{\Psi}_{q,-s} \} \qquad (29)$$

The ground state is defined as

$$|\Phi_0> = \prod_{|q_x|\leq q_\delta}[v_q + u_q \tilde{\Psi}^+_{q,s}\alpha_1\tilde{\Psi}_{q,s} + w_q \tilde{\Psi}^+_{q,s}\alpha_3\tilde{\Psi}_{q,s} + \theta_q c^+_{\bar{q},s}c^+_{q,s}c^+_{-\bar{q},-s}c^+_{-q,-s}]|\Phi_\delta>. \quad (30)$$

The relations between $v_q, u_q, w_q, \theta_q$ are given in [4], and the $\alpha$'s are the Dirac matrices. Notice that Eq. (30) does not use $\Psi_{q,-s}$ (in addition to $\Psi_{q,s}$), because the two spin states are already included in the definition of $\Psi_{q,s}$. The method used in [3,4] is that the unperturbed Hamiltonian is consistent with the condensed phase as given by the Hartree-Fock approximation. The perturbation Hamiltonian is used to calculate the Hartree-Fock condensation energies. The unperturbed Hamiltonian is given by

$$H_0 = \frac{1}{2}\sum_q[\tilde{\varepsilon}_{qx}\tilde{\Psi}^+_{q,s}\beta\tilde{\Psi}_{q,s} + \Lambda_q\tilde{\Psi}^+_{q,s}(x)\alpha_3\tilde{\Psi}_{q,s}(x) + \Delta_q\tilde{\Psi}^+_{q,s}\alpha_1\tilde{\Psi}_{q,s}] \quad (31)$$

The $\alpha_3$ component of the Hamiltonian includes terms like $c_{qs}c^+_{\bar{q}s}$ and $c_{-qs}c^+_{-\bar{q}s}$. Similar terms exist in the ground state. When one converts this operator representation into the spatial dependent state functions, one would get $f(x,G)\cos(2q_\delta x)$. With the addition of the time dependence, this is a standing CDW in the x-direction with the wavevector $2q_\delta$, which modulates the AFM SDW with the wavevector *G*. The modulation is such that the SDW switches between the AFM states C and D every cycle. As a result of the CDW, the momentum conservation may be violated by $2q_\delta$. Thus, in addition to the usual Greens' function

$$G_0(q,\omega) = \frac{\omega I + \Lambda_q\alpha_3 + \Delta_q\alpha_1 + \tilde{\varepsilon}_{qx}\beta}{\omega^2 - E_q^2 + i\delta}, \quad (32)$$

with $E_q^2 = \tilde{\varepsilon}_{qx}^2 + \Lambda_q^2 + \Delta_q^2$, one has the mixed momenta Greens' functions $G_0(\bar{q},q,\omega) = \alpha_0 G_0(q,\omega)$, and $G_0(q,\bar{q},\omega) = G_0(q,\omega)\alpha_0$.

The interaction Hamiltonian is given by

$$H_i = \frac{1}{8}\sum_{q,q',p}V_p(\tilde{\Psi}^+_{q'-p,s}\tau_3\tilde{\Psi}_{q',s})(\tilde{\Psi}^+_{q+p,s}\tau_3\tilde{\Psi}_{q,s}) - \frac{1}{2}\sum_q(\Lambda_q\tilde{\Psi}^+_{q,s}\alpha_3\tilde{\Psi}_{q,s} + \Delta_q\tilde{\Psi}^+_{q,s}\alpha_1\tilde{\Psi}_{q,s}). \quad (33)$$

The second term compensates for the same part in the unperturbed Hamiltonian, since the total Hamiltonian is $H = H_0 + H_i$. This second term is not applied in the calculations of the perturbation by means of the Feynman diagrams and Wick's theorem. For these calculations one should apply only the first term, which may be considered as a scattering process of the fields $\tilde{\Psi}_q$ and $\tilde{\Psi}_{q'}$ via the potential $V_p$, and the vertex $\tau_3$. Let us denote this first term by $H_i'$. The Hartree diagrams, such as those shown in Fig. (1) of Ref. [4], involve the Greens' functions $\alpha_0 G_0(q,\omega)$ (or $G_0(q,\omega)\alpha_0$), which suggests that the relevant vertex is $\alpha_3 = \tau_3 \alpha_0$. Consequently, all components of the Hartree diagrams, except for the $\alpha_3$ component, vanish. However, it is shown that a coherence field may be defined that is scattered via the vertices $\beta$, $\alpha_1$, and $\alpha_3$. Since these vertices scale with the same Dirac matrices as the Green's functions, the Hartree diagrams are finite and even large [3,4]. The coherence field is defined by

$$\Psi_{qs} = X_q \tilde{\Psi}_{qs}. \tag{34}$$

The transformation matrix is

$$X_q = -2w_q \alpha_3 - 2u_q \alpha_1 + (\theta_q - v_q)\beta \tag{35}$$

With the coherence field the two body part of the interaction Hamiltonian becomes

$$H_i = \frac{1}{8}\sum_{q,q',p} V_{\bar{p}}(\Psi^+_{q'-p,s} X^+_{q'-p} \alpha_3 X_{q'} \Psi_{q',s})(\Psi^+_{q+p,s} X^+_{q+p} \alpha_3 X_q \Psi_{q,s}), \tag{36}$$

with

$$X_q^+ \alpha_3 X_q = E_q^{-2}[(-E_q^2 + 2\Lambda_q^2)\alpha_3 + 2\Lambda_q \Delta_q \alpha_1 + 2\Lambda_q \tilde{\varepsilon}_{qx}\beta]. \tag{37}$$

Thus, with respect to the coherence field, the scattering vertex has the same matrix components as the Greens' function. Besides, the scattering potential which is used in the Hartree diagram is not of zero wavevector, but with the wavevector $2q_\delta$. It is argued in [2-4] that the total static interaction at zero momentum should vanish, but not so for the total interaction with momentum transfer of twice the Fermi wavevector. That interaction is significant. Therefore, we expect that all the three components of the Hartree diagram should result in significant contributions. These contributions are given by [4]

$$\Lambda_{q,H} = -U_H C N_{nes} \Lambda \frac{\Lambda_q^2 - \Delta_q^2 - \tilde{\varepsilon}_{qx}^2}{\Lambda_q^2 + \Delta_q^2 + \tilde{\varepsilon}_{qx}^2} \tag{38a}$$

$$\Delta_{q,H} = -U_H C N_{nes} \Lambda \frac{2\Lambda_q \Delta_q}{\Lambda_q^2 + \Delta_q^2 + \tilde{\varepsilon}_{qx}^2} \tag{38b}$$

$$\chi_{q,H} = -U_H C N_{nes} \Lambda \frac{2\Lambda_q \tilde{\varepsilon}_{qx}}{\Lambda_q^2 + \Delta_q^2 + \tilde{\varepsilon}_{qx}^2} \tag{38c}$$

In Eqs. (38) $\chi_{q,H}$ is the $\beta$ component of the Hartree self energy, $U_H$ is the Hartree potential, $\Lambda$ is the order parameter at the Fermi energy, $N_{nes}$ is the q density of states, and C is a constant of the order of 0.5 [3,4]. The relations between the Fock components and the Hartree components of $\Lambda_q$ and $\Delta_q$ were defined by $f_\Lambda$ and $f_\Delta$, respectively. The plausibility that $f_\Delta$ is negative was pointed out in Ref. [4], together with the criterion for the significance of the Hartree diagrams, which is $|U_H N_{nes} C| \geq 0.5$. The fulfillment of this criterion is plausible because $U_H$ carries a momentum transfer of $2q_\delta$, a momentum transfer at which the system is unstable.

## 5. CONCLUSIONS.

In the present work we have stressed the significance of the antibonding cellular function to the electronic correlations in the copper oxide HTSC. We have also shown that the AFM character of the parent undoped materials is related to the agglomeration of the charge stripes in the underdoped materials, through the exchange parameter *J*. The linear charge stripes were proven to be crucial for the creation of the AFM magnons that were observed in Neutron scattering experiments. Besides, this one dimensional vertical-horizontal characteristic of the holes is related to the horizontal- vertical segments in the density of states which, in turn, are responsible for the nesting in these directions. Consequently, the double correlations theory of Ref. [4] may be applicable in the anti-nodal directions of the $d_{x^2-y^2}$ state of the Copper, which makes the theory compatible with experiment in general, and with ARPES and Neutron scattering in particular. The importance of the Hartree diagrams has been stressed again. The Fock integrals alone (or some of their equivalent) cannot be compatible with the existence of both $\Lambda$ and $\Delta$, where $\Lambda > \omega_D > \Delta$, and $\omega_D$ is the Debye energy. Actually, the conjectured negative sign of $f_\Delta$ implies the frustration of the sign of $f_\Delta$, should the inapplicability of the Hartree diagram is assumed. This (among other considerations) confirms the relevance of the coherence field, and consequently, of the Hartree integral.

One should notice that the present analysis is still somewhat simplistic and idealistic (besides being a zero temperature treatment). The states $\Phi_q$ and $\Phi_{qx}$ are not the only states in which an injected hole can be. Our treatment does not include states in the nodal part of the zone (the round part of the Fermi arc). It does not include configurations with same spin states at some neighboring sites. One can show that a moving charge leaves such traces behind in its trail. We also did not discuss the nature of the excitations of the double correlated phase (apart from the discussion in [4], which is relevant only in part). Finally, the treatment applies only to the underdoped Copper oxide HTSC. Experiments show that with increasing doping, the straight nesting segments of the Fermi surface diminish together with the pseudogap. For overdoping, there is a range at which some nesting seems to be in the nodal directions, although its consequences are still unclear. All these issues need future considerations.

**FIGUE CAPTIONS**

**Fig. 1**

**1a.** A model illustrating how the neighboring stripes domains are related to each other.

**1b.** An illustration of how the states $\Phi_l^m$ are organized with respect to each other, and how the AFM states C and D are organized in relation to the column of holes. The pluses and minuses indicate the signs of the spin states, $s_z$. The shaded columns indicate stripes of holes.

**Fig. 2**

The wavevectors for which the uncondensed systems are unstable, are shown in the first reciprocal zone (in units of $a^{-1}$). The instability with respect to the central wavevector $a^{-1}(\pi,\pi)$ is responsible for creating the AFM state in undoped samples. The instabilities with respect to the satellites at $a^{-1}[\pi,\pi(1\pm\delta)]$, and $a^{-1}[\pi(1\pm\delta),\pi]$ are responsible for the stripes driven AFM magnons. The magnons' wavevectors modulate the AFM central wavevector. Both phenomena were observed by Neutron scattering experiments.

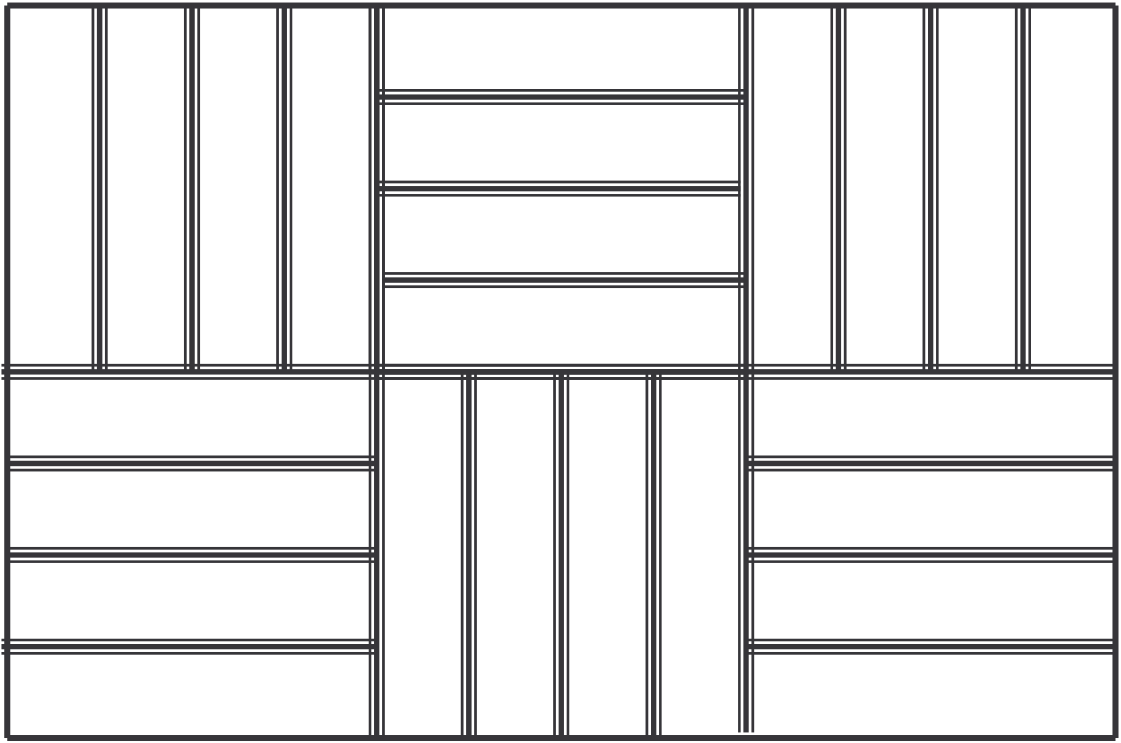

Fig. 1a

| | | | | | | | | | | | | | | | |
|---|---|---|---|---|---|---|---|---|---|---|---|---|---|---|---|
| | + | - | + | ---------- | - | + | - | | + | - | + | ---------- | + | - | + |
| | - | + | - | ---------- | + | - | + | | - | + | - | ---------- | - | + | - |
| | + | - | + | ---------- | - | + | - | | + | - | + | ---------- | + | - | + |
| | | | | | | | | | | | | | | | |
| | + | - | + | ---------- | - | + | - | | + | - | + | ---------- | + | - | + |
| | - | + | - | ---------- | + | - | + | | - | + | - | ---------- | - | + | - |

Fig. 1b

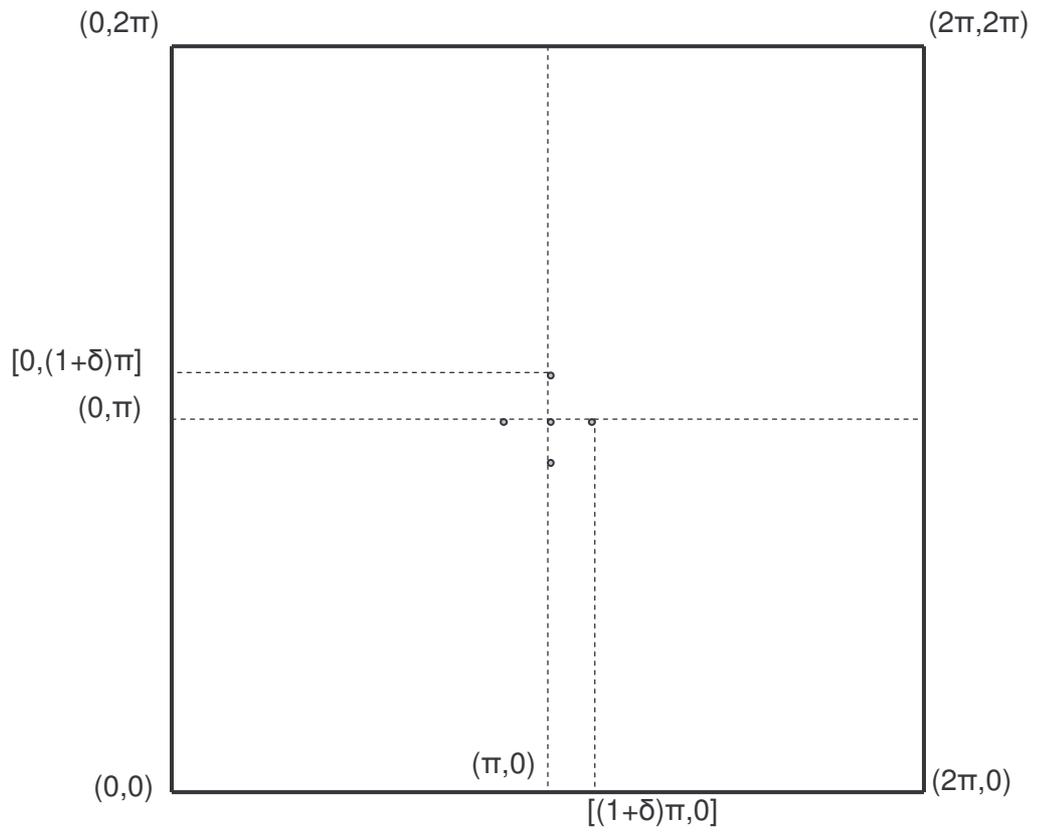

Fig. 2